\documentclass[12pt]{iopart}

\usepackage{graphicx}
\usepackage{hyperref}
\usepackage{float}

\begin{document}

\title{Spin supercurrent in Josephson contacts with noncollinear ferromagnets}
\author{Zahra Shomali$^1$, Malek Zareyan$^1$ and Wolfgang Belzig$^2$}

\address{$^1$ Institute for Advanced Studies in Basic Sciences (IASBS),
 P. O. Box 45195-1159, Zanjan 45195, Iran}
\address{$^2$ Fachbereich Physik, Universit\"at Konstanz, D-78457 Konstanz, Germany}


\begin{abstract}

  We present a theoretical study of the Josephson coupling of two
  superconductors which are connected through a diffusive contact
  consisting of noncollinear ferromagnetic domains. The leads are
  conventional s-wave superconductors with a phase difference of
  $\varphi$. First, we consider a contact with two domains
  with magnetization vectors misoriented by an angle $\theta$. Using
  the quantum circuit theory, we find that in addition to the charge
  supercurrent, which shows a $0-\pi$ transition relative to the angle
  $\theta$, a spin supercurrent with a spin polarization normal to the
  magnetization vectors flows between the domains. While the charge
  supercurrent, is odd in $\varphi$ and even in $\theta$, the spin
  supercurrent is even in $\varphi$ and odd in $\theta$. Furthermore,
  with asymmetric insulating barriers at the interfaces of the
  junction, the system may experience an
  antiferromagnetic-ferromagnetic phase transition for
  $\varphi=\pi$. Secondly, we discuss the spin supercurrent in an
  extended magnetic texture with multiple domainwalls. We find the
  position-dependent spin supercurrent. While the direction of the
  spin supercurrent is always perpendicular to the plane of the
  magnetization vectors, the magnitude of the spin supercurrent
  strongly depends on the phase difference between the superconductors
  and the number of domain walls. In particular, our results
  reveal a high sensitivity of the spin- and charge-transport in the junction to the number of domain
  walls in the ferromagnet. We show that superconductivity in
  coexistence with non-collinear magnetism, can be used in a Josephson
  nanodevice to create a controllable spin supercurrent acting as a
  spin transfer torque on a system. Our results demonstrate the
  possibility to couple the superconducting phase to the magnetization
  dynamics and, hence, constitutes a quantum interface e.g. between the
  magnetization and a superconducting qubit.

\end{abstract}

\maketitle

\section{Introduction}
The Josephson effect refers to the coherent transfer of Cooper pairs
between two weakly coupled superconductors \cite{Jos62}. In a
Josephson contact with a normal metal between the superconductors, the
underlying microscopic mechanism is Andreev scattering \cite{And64} at
the two normal-metal-superconductor interfaces which converts electron
and hole excitations of opposite spin directions into each other by
creating a Cooper pair. The resulting dissipationless electrical
current is driven by the difference between the phases of the
superconducting order parameters across the contact. From the fact
that superconductivity is a coherent state of spontaneously broken
U(1)-symmetry, it follows that the Josephson effect is a response of this coherent state to an inhomogeneity over the junction which is
produced by the variation of the phase. An analogous
non-superconducting effect is predicted to exist in the magnetic
tunnel barrier between two ferromagnets with the SO(3) symmetry
breaking coherent states \cite{Slo93,Nog04}. In this case the
misorientation angle $\theta$ of the two magnetization vectors is the
driving potential for a dissipationless spin supercurrent, similar
to the exchange interaction.
\begin{figure}
  \begin{center}
    \includegraphics[width=4.5in]{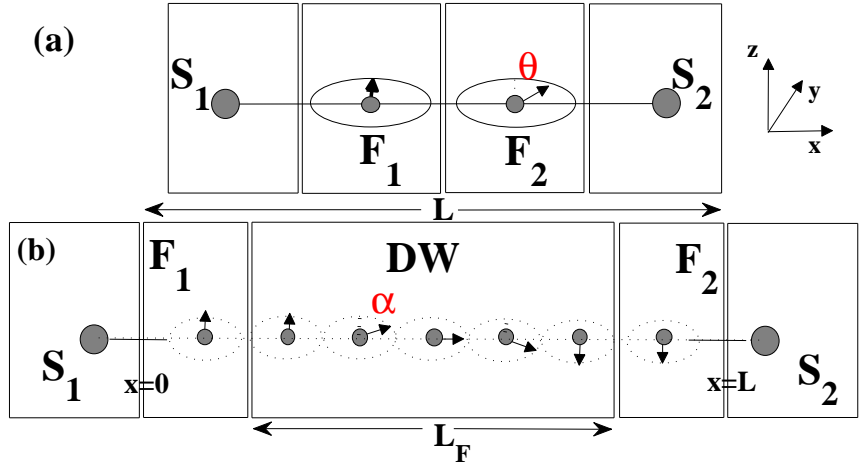}
  \end{center}
  \caption{\label{Fig:1}(a) (Color online) S$_1$F$_1$F$_2$S$_2$ junction, where, each
    ferromagnet is presented by a node. $\theta$ is the angle between
    the magnetization direction of the F$_1$ and F$_2$. (b) The S$_1$F$_1$DWF$_2$S$_2$
    junction with Neel domain wall (DW), $\alpha$ is the angle which
    the local exchange field in the domain wall makes with the
    y-direction.}
\end{figure}

\par
In this Article we develop the circuit theory of the
superconducting spin Josephson effect in an inhomogeneous
ferromagnetic (F) contact between two conventional
superconductors. We show that when the F contact consists of two
domains whose magnetization vectors enclose an angle $\theta$, in
addition to the charge supercurrent, a spin supercurrent will also
appear. This spin supercurrent is created by the simultaneous
existence of the superconducting states and a noncollinear
orientation $\theta$ of the magnetization vectors. Interestingly,
we find that the spin supercurrent and the corresponding spin
transfer torque is directed perpendicular to the plane of the two
magnetization vectors, such that it would lead to a precession of
the magnetizations around each other. In addition to extensive theoretical
\cite{Ber79,2Ber79,Bra06,Ral08} and experimental
\cite{Fre85,Hun88,Kla03,Tso03,San08} studies of the spin-transfer
torque in F spin-valve and domain structures, there have been
studies devoted to the spin-transfer torque in structures with
superconducting parts
\cite{Wai02,Lof05,Nus05,Zha08,Lin09,Gre09,Ali10,Gon07,Zha07,Zha08,Gia05,Gia07,Gia08,Hol11}.
The essential effect is the production of long-ranged spin-triplet
superconducting correlations by the interplay between the induced
spin-singlet correlations and the noncollinearity of the
magnetization profile in F contact
\cite{Ber01,2Ber01,Ber05,Bla04,Cro07,Bra07,Bra08}. Compared to previous studies, we present a quantum circuit theory calculation which takes the spatial variation of Green's functions as well as the nonlinearity of the proximity effect fully into account. By this method we specifically are able to obtain the inhomogeneity of the spin supercurrent and to define a spin-transfer torque in non-collinear ferromagnetic Josephson contacts.

Several experimental works on this triplet proximity effect have been
done \cite{Sos06,Kei06}. Recently, Khaire $et$ $al$. \cite{Kha10}
reported the observation of the long-range supercurrent in Josephson
junctions which is controllable by varying the thickness of one of the
ferromagnetic domains. Also, Robinson $et$ $al$. \cite{Rob10,2Rob10}
detected the flow of a long-range supercurrent in the ferromagnetic
Josephson junction with a magnetic Ho-Co-Ho trilayer and found an
enhancement of the critical currents in the antiparallel configuration
of the Junctions with a trilayer Fe/Cr/Fe barrier.

In analogy to the conventional charge Josephson effect, the spin
Josephson effect has the tendency to remove the inhomogeneity of the
order parameter vector of the spin-triplet superconducting state in
the F-contact. This is analogous to the spin Josephson effect in contacts
between two unconventional triplet superconductors, where the Cooper
pair spin current appears in conjunction with the usual charge
supercurrent\cite{Bry08,Bry09,2Bry09,3Bry09}.

The spin-dependent circuit theory has already been used in Ref. ~\cite{Bra07} to study the density of states, and the Josephson supercurrent in S/F/S heterostructures,  which are shown to be dependent on the configuration of the magnetization in F. Here, we further study the spin supercurrent and spin transfer torque in such Josephson junctions. We demonstrate the dependence of the charge and spin supercurrent
on the phase difference $\varphi$ and the angle between the
magnetizations $\theta$: the spin supercurrent is an even function of
$\varphi$ and an odd function of $\theta$, the charge supercurrent
satisfies the inverse relations relative to the $\varphi$ and
$\theta$.
Further, we study the equilibrium configuration of the exchange field vectors as a function of the phase difference and the temperature. We obtain phase diagrams which show the antiferromagnetic-ferromagnetic phase transitions in the system which hasn't been announced in any other literatures.

\par
We also discuss the generality of this effect for other
ferromagnetic contacts with a more complex inhomogeneity of the
direction of the magnetization vector. In particular, when the
ferromagnetic contact consists of an in-plane rotating magnetization
vector between two homogeneous domains with antiparallel
magnetization, we find that the spin supercurrent is highly
  sensitive to the value of the wave vector. We show that one can tune
  the spin supercurrent acting as a spin transfer torque, by changing
  the phase difference between the superconductors or with variations
  of the wave vector. Also, we investigate the position dependence of the spin current in S1F1DWF2S2 which, to the best of our knowledge, hasn't been studied in any other texts. We find that the behavior of spin supercurrent relative to the position strongly depends on the phase difference between the superconductors. We extend the quantum circuit theory by the description, how spin-transfer torques can be calculated within the method.

\section{Model and basic equation}

We describe the basic theory and the model first for a two-domain
ferromagnetic contact between two conventional superconductors, as is
shown schematically in Fig.~\ref{Fig:1}a.  The exchange field of one
domain F$_1$ makes an angle $\theta$ with that of the other domain
F$_2$. We restrict our study to the  time-independent case and do not
consider changes in the magnetic structure in this article. The generalization to the structure with a continuous
magnetization texture in Fig.~\ref{Fig:1}b is straightforward and
described in the end of this section.  To proceed our work, we make
use of the quantum circuit theory which is a finite-element technique
for calculating the quasiclassical Green's functions in diffusive
nanostructures \cite{Naz94,Naz99,bNaz05,Cot09,HH02,HH05}. In this
technique, we represent each  F domain and S reservoirs
by a single node, which is characterized by an energy-dependent
$4\times4$-matrix Green's function $\check{G}_{i}$, in Nambu and spin
($=\uparrow,\downarrow$) spaces \cite{HH02,HH05}. Furthermore, the two
nodes in the F domains are assumed to be weakly coupled to each other by
means of a tunneling contact. In terms of its spin-space matrix
components $\hat{g}$ and $\hat{f}$, the matrix Green's function is
written as
\begin{equation}
  \check{G}=\left(
    \begin{array}{cc}
      \hat{g} & \hat{f} \\
      \hat{f}^{\dag} & -\hat{g}
    \end{array}
  \right),
  \hat{a}=\left(
    \begin{array}{cc}
      a_{\uparrow\uparrow} & a_{\uparrow\downarrow} \\
     a_{\downarrow\uparrow} & a_{\downarrow\downarrow}
    \end{array}\right),a=f,g.
\end{equation}
We consider the equilibrium condition where a misorientation angle
$\theta$ and phase difference $\varphi$ may drive Josephson spin and
charge currents between two adjacent nodes.  These equilibrium
Josephson currents can be extracted from the matrix current defined as
\begin{equation}
\check{I}_{ij}=(g_{ij}/2)\left [\check{G}_{i},\check{G}_{j}\right],
\label{eq:matrixi}
\end{equation}
where $g_{ij}$ is the tunneling conductance of the contact between two
nodes and $i$ and $j$ denote the connected nodes. This approach works
also, if we divide the ferromagnetic region into $n$ nodes, as is
necessary in the case of Fig.~\ref{Fig:1}. Then we have to take
$(n-1)/g_{ij}=(1/g_{F1F2})-(1/g_{S1F1}+1/g_{S2F2})$. Where, the
conducting part of F domains is discretized into $n$ nodes. For a
S1F1F2S2 structure, $n$ is two. The conductance of the tunnel barrier
between S1(2) and F1(2) is denoted by $g_{S1(2)F1(2)}$ and, $g_{F1F2}$
is the conductance of the whole F1F2 contact.
\par
The matrix current obeys the following law of current conservation
in matrix form
\begin{equation}
  \check I_{\omega i}+\check I_{s i}+\sum_{j=i\pm1}\check I_{ij}=0\,.
  \label{eq:balance}
\end{equation}
Here $\check I_{\omega i}=- G_{Q}(\omega
/\delta_{i})\left[\check\tau_{3},\check G_i\right]$, with $\delta_i$
being the electronic level spacing of the node, is the matrix of the
leakage current which takes into account the dephasing of electrons
and holes due to their finite dwell time in the node $i$, and $\check
I_{s i}= i(G_{Q}/\delta_{i})
\left[(h^{x}_{i}\hat{\sigma}_{1}+h^{y}_{i}\hat{\sigma}_{2}+h^{z}_{i}\check{\tau}_{3}
  \hat{\sigma}_{3}),\check{G}_{i}\right]$ is the corresponding matrix
current representing the leakage caused by the spin-splitting due to
an exchange field $\vec{h}_i$ ($G_{Q}\equiv e^{2}/(2\pi\hbar)$ is the
quantum of the conductance). The third term represents the
matrix currents from the neighboring nodes i-1,i+1. From Eq.~(\ref{eq:balance}) we can find the spin-torque, e.g. in z-direction as $\tau_{zi}=I_{zi,i+1}+I_{zi,i-1}$.

Equation (\ref{eq:balance}) is given for all nodes and is supplemented
by boundary conditions, which are the values of $\check{G}$ in the S
reservoirs. We neglect the inverse proximity effect in the reservoirs
and set the matrix Green's function in S1 and S2 to the bulk values:
\begin{equation}
  \check{G}_{1,2}=\frac{\omega\check\tau_3+{\check\Delta}_{1,2}}{\sqrt{\omega^{2}+\left|\Delta\right|^2}}
\end{equation}
where
\begin{equation}
{\check{\Delta}}_{1,2}=\left(
         \begin{array}{cc}
           0 & {\hat\Delta}_{1,2} \\
           {\hat\Delta^{\dagger}}_{1,2} & 0 \\
         \end{array}
       \right)\,,\,
  \check\tau_3=\left(
    \begin{array}{cc}
      \hat{1} & \hat{0} \\
      \hat{0} & -\hat{1} \\
    \end{array}
  \right).
\end{equation}
Here, ${\hat\Delta}_{1,2}=|\Delta|\exp{(\pm i\varphi/2)}\hat\sigma_1$
are, respectively, the superconducting order parameter matrix in S1
and S2 ($\hat\sigma_i$ denote the Pauli matrices in spin space) and
$\omega=\pi T(2m+1)$, with $m$ being an integer, is the Matsubara
frequency. The temperature dependence of the amplitude of the order
parameter is well approximated by $|\Delta|$$=1.76T_{c}
\textrm{tanh}(1.74\sqrt{T_c/T-1})$. We note that the matrix Green's
function satisfies the normalization condition
$\check{G}^{2}=\check{1}$.

We have solved these equations numerically by an iteration method. In
our calculation we start by choosing a trial form of the matrix
Green's functions of the nodes, for a given $\phi$, T, and the
Matsubara frequency m=0. Then, using Eq. (\ref{eq:balance}) and the
boundary conditions iteratively, we refine the initial values until
the Green's functions are calculated in each of two nodes with the
desired accuracy. Note that in general for any phase difference
$\phi$, the resulting Green's functions vary from one node to another,
simulating the spatial variation along the F contact (see
Ref.~\cite{Shomali}). From the resulting Green's functions and
Eq. (\ref{eq:matrixi}) we find the matrix currents. Then, we calculate
the charge supercurrent $I$ and the components of the spin
supercurrent vector ${\vec I}$ from the relations
\begin{equation}
  \label{eq:currents}
  I_{ij}=\textrm{tr}\hat\sigma_3\check I_{ij} \,,\,
  I_{zi,j}=\textrm{tr}\check\tau_3\hat\sigma_3\check I_{ij} \,,\,
  I_{x(y)i,j}=\textrm{tr}\hat\sigma_{1(2)}\check I_{ij}\,,
\end{equation}
in which $\textrm{tr}\ldots=(i\pi T/2e) \sum_\omega\textrm{Tr}\ldots$
with $\textrm{Tr}$ denoting the trace in Nambu-spin spaces. In the
next steps we change to the next Matsubara frequency and use the
results from the previous one $\omega_{m-1}$ as initial guess. We find
the respective contribution  to the spectral currents and continue to higher
frequencies until the required precision of the summation over m is
achieved.

In the following we scale the length of the system, L, in units of the
diffusive superconducting coherence length $\xi$, and use the
dimensionless parameters of $h/T_{c}$ and $t=T/T_{c}$, as measures of
the amplitude of the exchange field $h$ and the temperature $T$.To describe a continuous domain wall, we use as parameter the wave vector $Q$ associated with one full winding of the magnetization by $2\pi$. Hence the total number of windings of the magnetization is given by $QL_F/2\pi$, where $L_{F}$ is the length of the inhomogeneous region (see Fig.~\ref{Fig:1}). The
currents are expressed in units of $I_{0}$, the amplitude of the
critical Josephson charge current at $h=0$ and $t=0$.

\section{Results and discussions}

From the numerical calculations we obtain the spin supercurrent
with a polarization directed normal to the plane of the magnetization
vectors. Further, we find a transition of the favorable configuration
of the domain, from antiparallel to the parallel as the exchange field
of the asymmetric domains increases. Also, we show that in a system
with more complex configuration of the direction of the magnetization,
the profile and penetration depth of the spin supercurrent are highly
dependent on the number of the rotations that the magnetization vector
has undergone across the domain wall.

\begin{figure*}
  \begin{center}
  \includegraphics[width=6in]{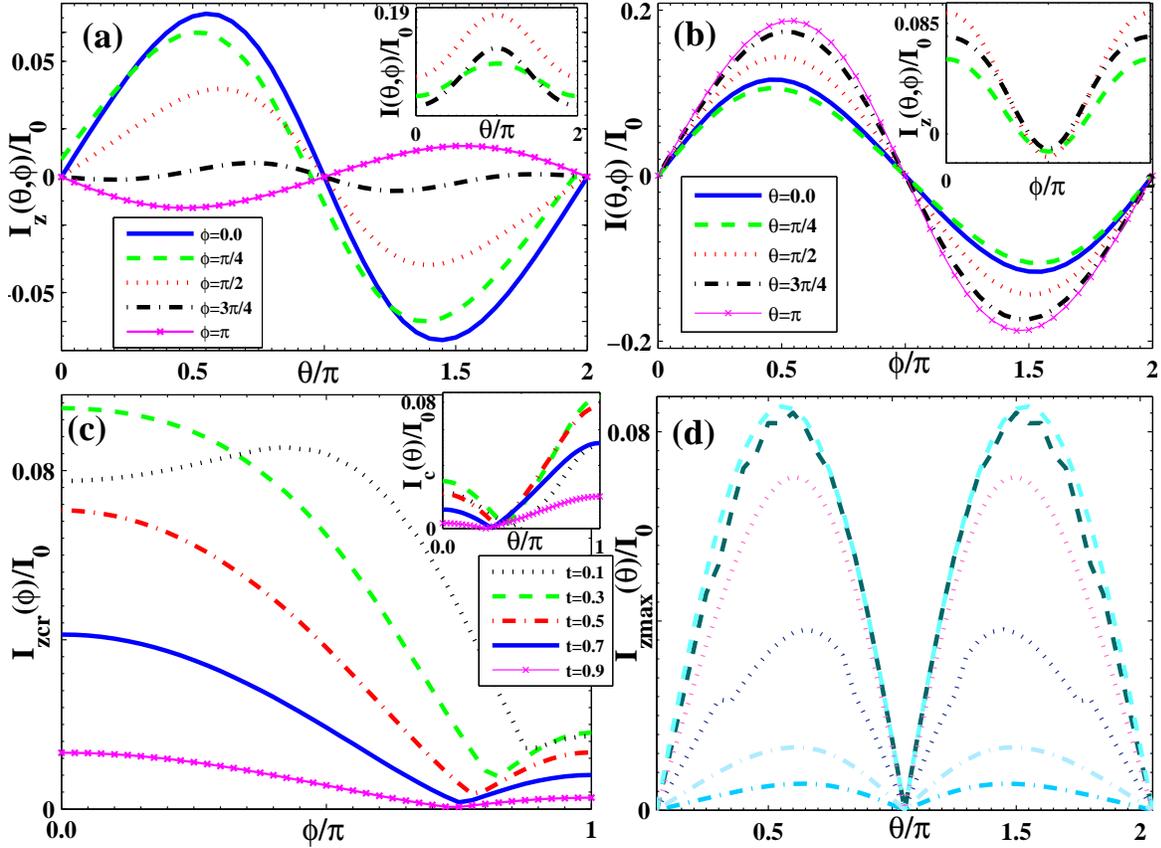}
  \end{center}
  \caption{\label{Fig:2}(a) (Color online) Plot of spin supercurrent
    versus $\theta$ for different values of $\varphi$ when
    $L/\xi=1.0$, $h/T_c=5.0$, $t=0.5$. (inset) $\theta$ dependence of
    charge supercurrent for the same system. (b) Plot of $I$ versus
    $\varphi$ for different values of $\theta$ for the previous
    system. (inset) $\varphi$ dependence of spin supercurrent for the
    same system. (c) Critical spin supercurrent versus $\phi$ for
    different temperatures, which shows the appearance of $0-\pi$
    transition relative to the $\phi$. (inset) Critical charge
    supercurrent versus $\theta$. (d) Light dashed, dotted,
    dashed-dotted lines are, respectively, the maximum spin
    supercurrent when $\varphi$ changes between $0$ and $\pi$ versus
    $\theta$, when $t=0.1,0.5,$ and $0.9$. Dark dashed, dotted,
    dashed-dotted lines are, respectively, spin supercurrent for the
    $\varphi$ which maximize the charge supercurrent for different
    temperatures for the same situation.}
\end{figure*}

\subsection{S1F1F2S2 junction}
Using the method described in Sec. 2 we have calculated the spin and
charge supercurrents for the two domains F contact of
Fig. \ref{Fig:1}a, when the exchange field vectors are taken to be in
the $x-y$ plane. We found that the spin supercurrent has a
polarization which is aligned along the $z$ axis, namely perpendicular
to the plane of the exchange fields of F$_1$ and F$_2$. Our results
for the dependence of the spin $I_{z}$ and charge $I$ supercurrents on
the misorientation angle $\theta$, the phase difference $\varphi$ and
the temperature $t$ are shown in Fig.~\ref{Fig:2}. We found that, in
general, the spin supercurrent obeys the symmetry relations
$I_{z}(\varphi)=I_{z}(-\varphi)$ and $I_{z}(\theta)=-I_{z}(-\theta)$,
which are the analogs of the relations $I(\varphi)=-I(-\varphi)$ and
$I(\theta)=I(-\theta)$ for the charge supercurrent (see
Fig.~\ref{Fig:2} a,b). These behaviors suggest that one can change the
direction of the spin supercurrent, which is proportional to the
induced spin transfer torque \cite{Bra06}, by changing the phase
difference between two superconductors when $\theta$ is fixed. We note
that a nonzero spin supercurrent is provided by a noncollinear
orientation of the exchange field vectors and the existence of the
superconductivity ($|\Delta|\neq 0$), even for $\varphi=0$.

We define the critical spin supercurrent, $I_{z cr} (\varphi)$, as the
maximum of the absolute value of the spin supercurrent as a function
of $\theta$ for a given $\varphi$, in similarity to the definition of
the charge critical supercurrent. We may also use a distinct
definition, which we denote by $I_{z max}(\theta)$, as the absolute
value of the spin supercurrent for a value of $\varphi$ which
maximizes the charge supercurrent as a function of $\varphi$, for a
given $\theta$. Figure.~\ref{Fig:2}c shows the behavior of $I_{z cr}$
as a function of $\varphi$ for different temperatures. At a given
temperature $t$, $I_{z cr}(\varphi)$ shows a change of sign at a phase
difference which depends on $t$. This change of sign may be recognized
as the signature of a transition between $0$ and $\pi$ spin Josephson
couplings, in analogy to the charge $0-\pi$ transition in F Josephson
junctions \cite{Bul77,Buz82,Buz91,Buz92,Rya01,Cht01,Kon02,Gui03,Buz05,Vas08}.  The corresponding critical charge
current shows a $0-\pi$ transition with varying $\theta$. Note that both
$I_{z cr}(\varphi)$ and $I_{cr}(\theta)$ have a nonzero value at the
transition point at low temperatures, as the signature of nonzero
second harmonic in the current-phase and the current-angle relations
\cite{Sel03,Moh06,Kon08}. In Fig.~\ref{Fig:2}d we have also plotted
$I_{z max} (\theta)$, which shows a change of sign at $\theta=\pi$ for
all temperatures \cite{Ali10}.

\begin{figure}
  \begin{center}
    \includegraphics[width=4in]{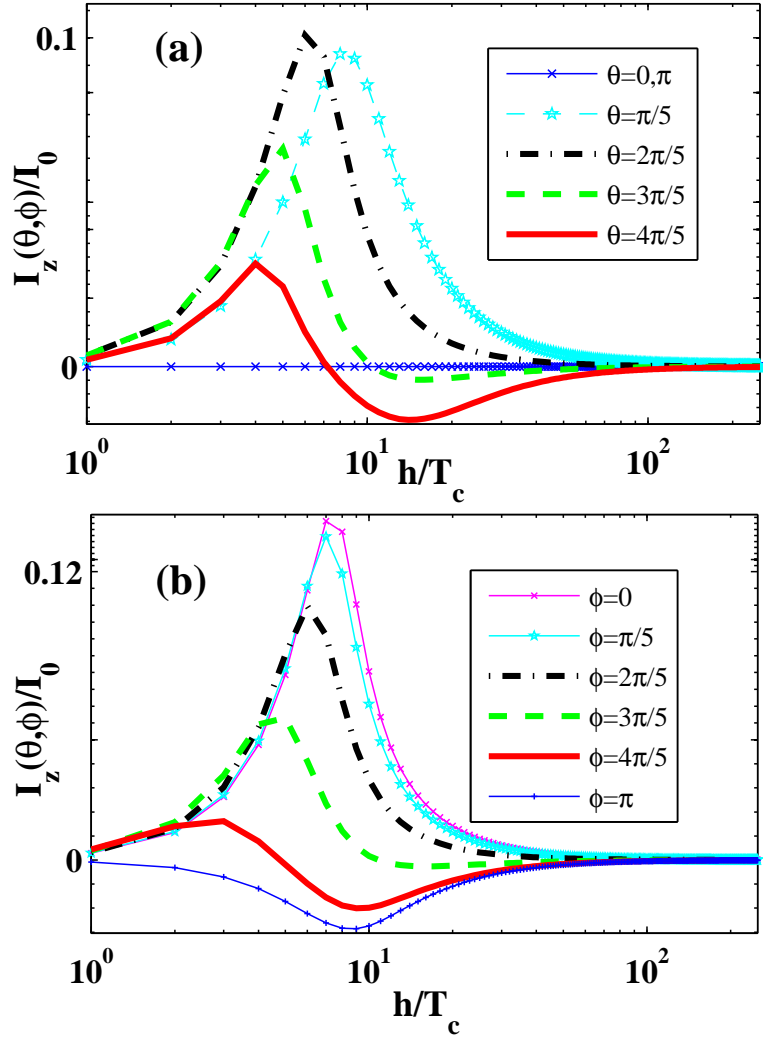}
  \end{center}
  \begin{center}
    \caption{\label{Fig:3} (Color online)(a) $I_{z}$ versus $h/T_c$
      for different values of $\theta$ when $\varphi=\pi/2$,
      $L/\xi=1.0$, and, $t=0.1$. (b) $I_{z}$ versus $h/T_c$ for the
      same system but for different $\varphi$ when $\theta=\pi/2$. (a)
      and (b) are logarithmic plots which show the $0-\pi$ transition
      relative to the exchange field.}
    \end{center}
\end{figure}

We have also studied the dependence of the spin supercurrent on the
absolute value of the exchange field. The results are shown in
Fig.~\ref{Fig:3}, in which $I_{z cr}$ is plotted as a function of
$h/T_c$ for different $\varphi$ and $\theta$. These results show that
the sign and the amplitude of the spin supercurrent can be also
modulated by varying $h/T_c$, which can be used for further tuning of
the corresponding spin transfer torque. We note that for strong ferromagnets with $h\gg \Delta$, the spin supercurrent vanishes.
This is due to the suppression of the amplitude of the Andreev reflection at S$_1$F$_1$ and S$_2$F$_2$ interfaces in this limit, which
suppresses the proximity effect.

\begin{figure}
\begin{center}
\includegraphics[width=5in]{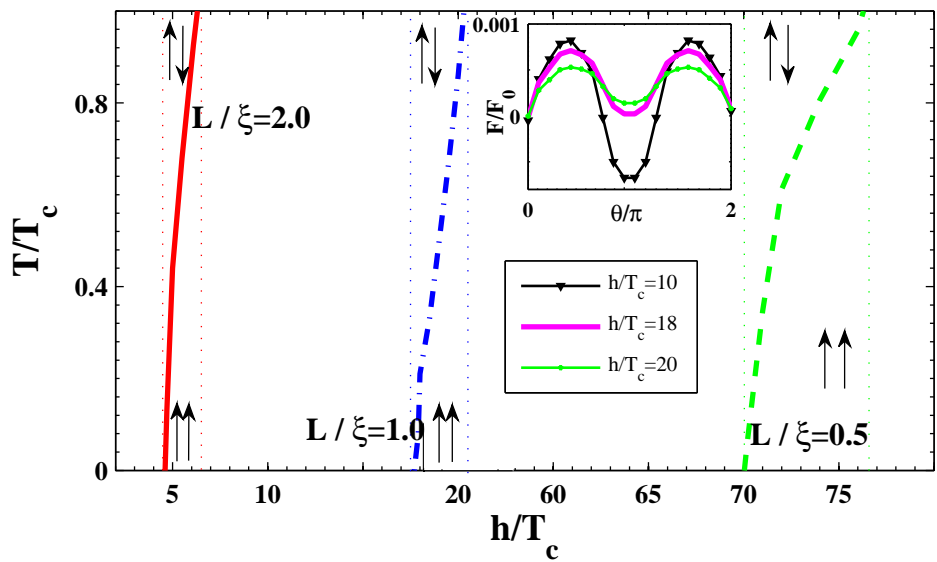}
\end{center}
\caption{\label{Fig5} (Color online) Phase diagram of the transition
  of the minimum of free energy from $\theta=\pi$ to $\theta=0$ for
  asymmetric systems with $g_{S1F1}=0.1 g_{S2F2}$ and different
  lengths. The inset shows the $F-\theta$ plane of the 3-dimensional
  energy plot when $L/\xi=1$ and $t=0.4$. Here, $F_{0}$ is the free
  energy of the system when the magnetization is collinear.}
\end{figure}

It is also interesting to study the equilibrium configuration of the
exchange field vectors as a function of the phase difference and the
temperature. The equilibrium angle can be obtained by minimizing the
free energy, $F$, of the contact as a function of $\theta$. We have
calculated the $\theta$-dependence of $F$ by integrating the spin
supercurrent over $\theta$ and charge supercurrent over $\phi$:
\begin{equation}
  \label{eq:ftfi}
  F(\varphi,\theta)= \int_{0}^{\varphi}I({\varphi}^{'},0)d{\varphi}^{'}+\int_{0}^{\theta}I_z(\varphi,{\theta}^{'})d{\theta}^{'}.
\end{equation}
Our calculation shows that the exchange field vectors favor
either parallel ($\theta=0$) or antiparallel ($\theta=\pi$)
configurations, depending on $\varphi$, $t$ and $h/T_c$. The behavior
of this superconductivity-induced exchange coupling differs for the two
cases of a contact with symmetric barriers with $g_{S1F1}=g_{S2F2}$ and an
asymmetric contact with a very different $g_{S1F1}$ and $g_{S2F2}$. For a
symmetric system, we have found that the coupling is antiferromagnetic
($\theta=\pi$) for $\varphi=0$, but becomes ferromagnetic ($\theta=0$)
for $\varphi=\pi$. This behavior is found to hold irrespective of the
values of $L/\xi$, $h/T_c$.

However, for an asymmetric system it is possible to change the
coupling from ferromagnetic to antiferromagnetic and vice versa by
varying $L/\xi$, $h/T_c$, or $t$, for the phase difference
$\varphi=\pi$. In Fig.~\ref{Fig5}, we have shown the
ferromagnetic-antiferromagnetic coupling phase diagram of the system
in the plane of $h/T_c$ and $t$, when $\varphi=\pi$ and for some
different values of $L/\xi$. This phase diagram is similar to the
$0-\pi$ Josephson couplings phase diagram of a homogenous F contact
between two superconductors, see Ref.~\cite{Shomali}. As we show in the
inset of Fig.~\ref{Fig5}, the minimum of $F$ as a function of $\theta$
shifts from $\theta=\pi$ for low values of $h/T_c$ to $\theta=0$ at
higher $h/T_c$s, when $L/\xi=1$ and $t=0.4$. The temperature induced
transition between the ferromagnetic and the antiferromagnetic phases
is also possible but only over a finite interval $\triangle h$ of the
amplitude of the exchange field of the F domains. This width of the
temperature induced transition increases with decreasing $L/\xi$.  For
$\varphi=0$, the coupling between the exchange fields of the two
domains is found to be always antiferromagnetic in an asymmetric
structure, which is very similar to the symmetric case.

\subsection{S1F1DWF2S2 junction with Neel domain wall}

The superconducting spin Josephson effect described above may take
place in F contacts with a more complex profile of the exchange field
vector. An interesting case is a finite width F domain wall between
two domains, where the exchange field has a continuous spatial
rotation between two homogeneous F domains. Here, we present the
results of our calculation of the spin supercurrent and the spin
transfer torque for a Neel domain wall junction, which is shown
schematically in Fig.~\ref{Fig:1}b. Note that the spin transfer torque
acting on the local magnetization on node $i$ is obtained from $\tau_{zi}=I_{zi,i+1}+I_{zi,i-1}$, as we have shown earlier. We model the
local angle of the exchange field vector with respect to the $y$-axis
to vary as $\alpha(x)=(QL_F)(x/L_F )$.

\begin{figure}
\begin{center}
\includegraphics[width=4in]{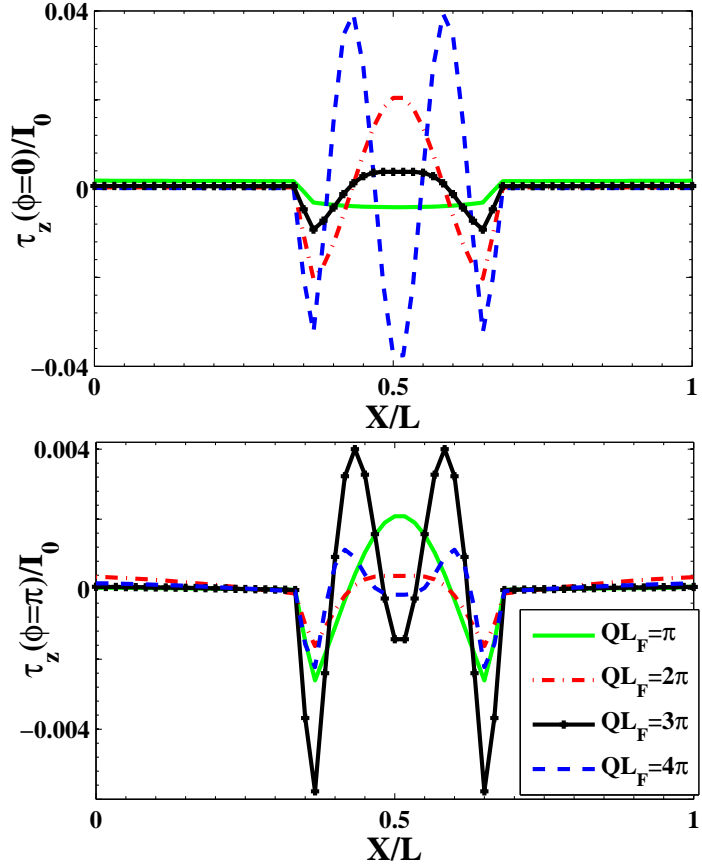}
\end{center}
\caption{\label{Fig:5} (Color online) Plot of $I_{z}$ versus position for the
  system with $L/\xi=1.0$, $L_F/L=1/3$, $h/T_c=5.0$, $\varphi=0.0$, $t=0.1$, and
  $QL_F$=$\pi$; $2\pi$; $3\pi$ and $4\pi$. Solid, dashed, dotted,
  and dashed-dotted lines, respectively, represent the results for
  $QL_F$=$\pi$; $2\pi$; $3\pi$ and $4\pi$. (b) The same as (a) but
  for $\varphi=\pi$.}
\end{figure}

We have obtained the position-dependent spin supercurrent, which is
perpendicular to the plane of the magnetization and flows in the
homogeneously and inhomogeneously magnetized parts of the system. To
the best of our knowledge, there have been no previous studies
investigating the position dependence of the spin current and the spin
transfer torque. Figures~\ref{Fig:5}(a,b) show the behavior of the
spin transfer torque versus position for a system with $L_F/L=1/3$. We
can see that the system has an interesting behavior depending strongly
on the values of $QL_F$. As we see in Fig.~\ref{Fig:5}a, the spin
transfer torque penetrating into the homogeneous ferromagnets becomes of negligible constant value. This shows that the spin supercurrent in F$_1$ and F$_2$ has a linear position dependence when $\phi=0$. In particular, this is true while the value of the spin supercurrent is comparable to the one in the nonhomogeneous parts.

 Also Fig.~\ref{Fig:5}b shows that the penetrating spin transfer torque in the homogeneous parts is nearly zero for $QL_F=\pi,3\pi$ and is much smaller than the one in the domain wall region for $QL_F=2\pi,4\pi$ when $\phi=\pi$. In addition, we note that the the spin transfer torque has
always a symmetric position dependence around $x=L/2$, whereas the
spin supercurrent always shows an asymmetric behavior. Our
calculations demonstrate further that the behavior of the spin current
and spin transfer torque versus position for a system with
$QL_F\pi$ and $\phi=\pi$ is similar to that of the system with
$QL_F=(n+1)\pi$ and $\phi=0$, see Fig.~\ref{Fig:5}. This observation
of a symmetry between the magnetic winding number and the
superconducting phase is presently not fully understood, but will be
the subject of future research.

\begin{figure}
\begin{center}
\includegraphics[width=4in]{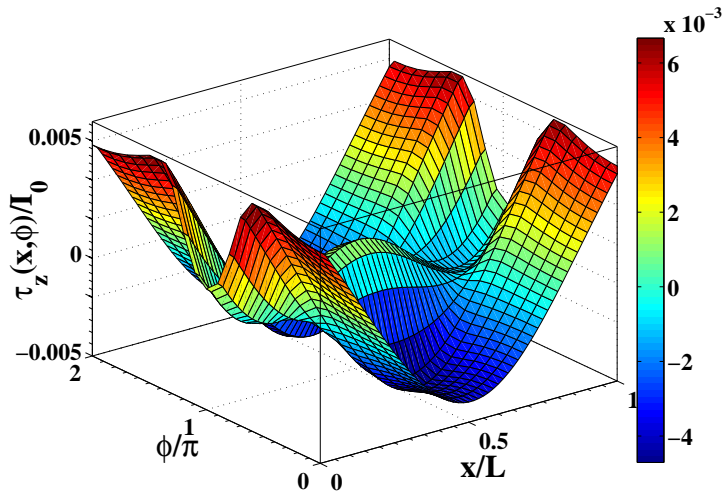}
\end{center}
\caption{\label{Fig:6} (Color online) Spin transfer torque $I_{z}$ versus
  position and phase difference between the superconductors for the
  Josephson system without any homogeneous ferromagnet, $L_F/L=1$,
  while $L/\xi=1.0$, $h/T_c=5.0$, $t=0.5$, and $QL_F$=$\pi$.}
\end{figure}

Finally, we have also studied Josephson systems without the
homogeneous ferromagnetic parts F$_1$ and F$_2$, which corresponds to
$L_F=L$. In this way we would like to check how the spin currents in
the homogenous part, which are appreciable in size, but give rise to a
negligible spin-transfer torque, influence the spin-torque on the
magnetization texture.  We show the corresponding dependence of the
spin supercurrent on the phase difference and the position in
Fig.~\ref{Fig:6} for a wall with rotation angle $\pi$. The first thing
to note, is that we observe a qualitatively similar phase dependence
as in the case with two homogeneous ferromagnets. In particular, the
$I_z(\phi)$ relation is always a symmetric. Also, for a specific
position, the sign of the spin transfer torque can be changed by
varying the phase difference between the superconductors. In addition,
for a simple Neel domain wall, when $QL_F=\pi$, we see that the
behavior of the spin transfer torque can be approximately described by
the relation $cos[(x/L)(QL_F)+\pi/2]$ for $\phi=0$, which turns
into $cos[3(x/L)(QL_F)+\pi/2]$ for $\phi=\pi$. These findings
have the same symmetry, as we mentioned in the previous
discussion for a system with homogenous ferromagnet attached. Also,
this result shows the strong dependence of the spin transfer torque on
the phase difference between the superconductors, which can be
interpreted as a direct interplay between the induced spin-singlet
correlations of the superconductors and the inhomogeneity of the
magnetization in the domain wall. In fact this interplay leads to the
generation of the triplet correlations in the contact region, whose
inhomogeneity drives the spin supercurrent and therefore the
spin-transfer torque.


\begin{figure}
  \begin{center}
    \includegraphics[width=5in]{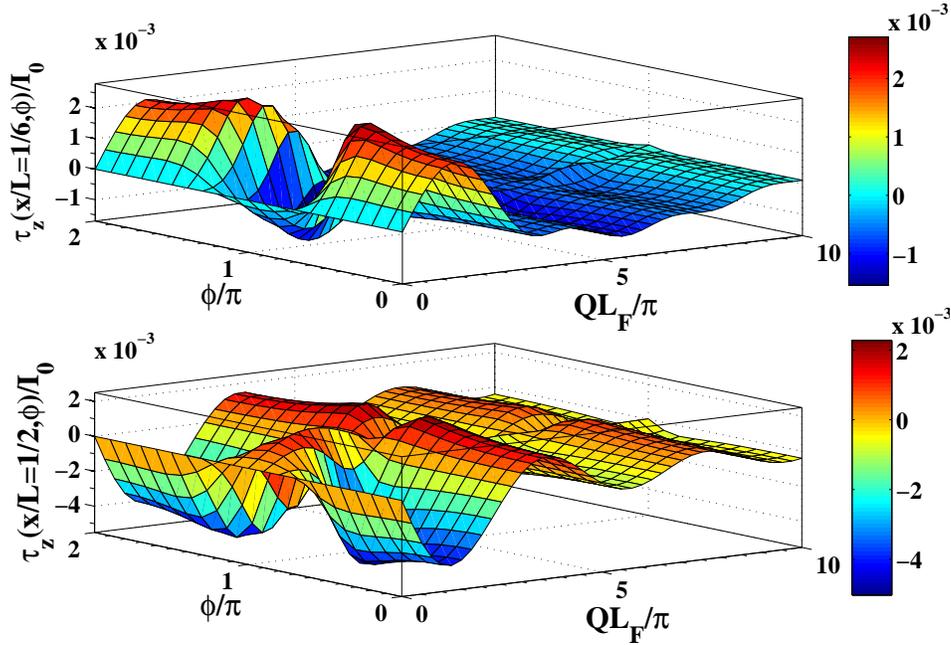}
  \end{center}
  \begin{center}
  \caption{\label{Fig:7} (Color online)(a) Spin transfer torque of $I_{z}$ versus $QL_F$ and $\phi$ for the
  system with $x/L=1/6$,$L/\xi=1.0$, $L_F/L=1.0$, $h/T_c=5.0$ and $t=0.1$. (b) Same as the previous one but
  for $x/L=1/2$. }
    \end{center}
\end{figure}

Figures~\ref{Fig:7}(a,b) show the spin transfer torque as function
of the phase difference, $\phi$, and the wave vector, $QL_F$ (which is
more or less given by the number of 180 degree domain walls). In
Figs.~\ref{Fig:7}(a,b) the behavior for different positions $x/L=1/2$ and
$x=L/6$ are shown. For the position closer to the superconductor, $x/L=1/6$, the spin transfer torque goes to
zero much faster for larger wave vectors than in the middle of the
ferromagnet at $x=L/2$. Hence, we see that the suppression of the spin
transfer torque strongly depends on the position in the domain wall.
Also, Fig.~\ref{Fig:7}(a,b) shows that the spin transfer torque
oscillations versus wave vector strongly depend on the phase difference
and also on position. While they start for small $QL_F$ in exactly
opposite fashion for both positions, the oscillations in the middle of
the domain wall are well behaved also for large $QL_F$, whereas the
behavior is more complex in the ferromagnetic part close to the
superconductor. In that limit no well defined oscillation period can be
identified. Furthermore, while the direction of the spin supercurrent
is fixed, we find that the sign and magnitude of the spin supercurrent
and spin transfer torque can be modulated by changing the wave vector
or the phase difference. We have obtained that superconductivity in
coexistence with non-collinear magnetism can be used in a Josephson
nanodevice to create the tunable spin supercurrent which acts as a
spin transfer torque on the junctions magnetization.

\section{Conclusion}

We have studied the Josephson effect in a diffusive contact consisting
of two ferromagnetic domains with noncollinear magnetizations which
connects two conventional superconductors. Using quantum circuit
theory, we have shown that the spin supercurrent will flow through the
contact, due to the generation of an inhomogeneous spin-triplet
superconducting correlations. The polarization of the spin
supercurrent is directed normal to the plane of two magnetization
vectors in a way that the resulting spin transfer torque intends to
align the magnetization of the two domains. The spin and charge
current-phase-angle relations obey the specific symmetry relations
versus the phase difference $\varphi$ and the misorientation angle
$\theta$. Whereas, the charge supercurrent satisfies the odd-even
relationship on the $\varphi$ and $\theta$, the spin supercurrent is
an even-odd function of $\phi$ and $\theta$. From these relations we
have predicted a transition between $0$ and $\pi$ Josephson coupling
by varying the misorientation angle $\theta$. We found a transition of
the favorable configuration of the domain, from antiparallel to the
parallel as the exchange field of the domains increases. This
transition occurs for asymmetric systems with $\phi=\pi$. Also, the
domains in the symmetric systems settle in a parallel configuration
when $\phi=\pi$.

We have further discussed the generation of the spin supercurrent in
magnetic contacts with more complex configuration of the direction of
the magnetization vector. For a domain wall between two domains with
antiparallel magnetizations, we have shown that the profile and
penetration depth of the spin supercurrent are highly dependent on the
number of the rotations that the magnetization vector has undergone
across the domain wall. In particular, we show that, while the direction of
the spin supercurrent is always perpendicular to the plane of the
magnetization vectors, the sign and the magnitude of the spin
supercurrent strongly depends on the phase difference between the
superconductors and the value of the wave vector. We present a
Josephson nanodevice which can be used to create a controllable spin
supercurrent and spin transfer torque.

\section{Acknowledgments}
We acknowledge financial support from the DFG through SFB 767 and from
the Research Initiative \textit{UltraQuantum}. Also, Z. S. and
M. Z. would like to thank W.B. for the financial support and
hospitality during their visit to University of Konstanz.


\section*{References}
\begin{thebibliography}{10}

\bibitem{Jos62} Josephson B D 1962 {\it Phys. Lett.} {\bf 1} 251
\bibitem{And64} Andreev A F 1964 {\it Sov. Phys. JETP} {\bf 19} 1228
\bibitem{Slo93} Slonczewski J C 1993 {\it J. Magn. Magn. Mater.} {\bf 126} 374
\bibitem{Nog04} Nogueira F C and Bennemann K H 2004 {\it Europhys. Lett.} {\bf 67} 620
\bibitem{Ber79} Berger L 1979 {\it J. Appl. Phys.} {\bf 3} 2156
\bibitem{2Ber79} Berger L 1979 {\it J. Appl. Phys.} {\bf 3} 2137
\bibitem{Bra06} Brataas A, Bauer G E W, and  Kelly P 2006 {\it Phys. Rep.} {\bf 427} 157

\bibitem{Ral08} Ralph D C and Stiles M D 2008 {\it J. Magn. Magn. Mater.} {\bf 320} 1190
\bibitem{Fre85} Freitas P P and Berger L 1985 {\it J. Appl. Phys.} {\bf 57} 1266
\bibitem{Hun88} Hung C Y and Berger L 1988 {\it J. Appl. Phys.} {\bf 63} 4276
\bibitem{Kla03} Kl\"{a}ui M, Vaz C A F, Bland J A C, Wernsdorfer W, Faini G, Cambril E and Heyderman L J 2003 {\it Appl. Phys. Lett.} {\bf 83} 105
\bibitem{Tso03} Tsoi M, Fontana R E and Parkin S S P 2003 {\it Appl. Phys. Lett.} {\bf 83} 2617
\bibitem{San08} Sankey J C, Cui Y -T, Sun J Z, Slonczewski J C, Buhrman R A and Ralph D C 2008 {\it Nat. Phys.} {\bf 4} 67
\bibitem{Wai02} Waintal X and Brouwer P W 2002 {\it Phys. Rev.} B {\bf 65} 054407
\bibitem{Lof05} L\"ofwander T, Champel T, Durst J and Eschrig M 2005 {\it Phys. Rev. Lett.} {\bf 95} 187003
\bibitem{Nus05} Nussinov Z, Shnirman A, Arovas D P, Balatsky A V and Zhu J -X 2005 {\it Phys. Rev.} B {\bf 71} 214520
\bibitem{Lin09} Linder J, Yokoyama T and Sudb\"{o} A 2008 {\it Phys. Rev.} B {\bf 79} 224504
\bibitem{Gre09} Grein R, Eschrig M, Metalidis G and Sch\"on G 2009 {\it Phys. Rev. Lett.} {\bf 102} 227005 (2009)
\bibitem{Ali10} Alidoust M, Linder J, Rashedi GH, T. Yokoyama and Sudb\"{o} A 2010 {\it Phys. Rev.} B {\bf 81} 014512
\bibitem{Gon07} González E M, Folgueras A D, Escudero R, Ferrer J, Guinea F and Vicent J L 2007 {\it New Journal of Physics} {\bf 9} 34
\bibitem{Zha07} Zhao E and Sauls J A 2007 {\it Phys. Rev. Lett.} {\bf 98} 206601
\bibitem{Zha08} Zhao E and Sauls J A 2008 {\it Phys. Rev.} B {\bf 78} 174511
\bibitem{Gia05} Giazotto F, Taddei F, Fazio R and Beltram F 2005 {\it Phys. Rev. Lett.} {\bf 95} 066804
\bibitem{Gia07} Giazotto F, Taddei F, D'Amico P, Fazio R and Beltram F 2007 {\it Phys. Rev.} B {\bf 76} 184518
\bibitem{Gia08} Giazotto F and Taddei F 2008 {\it Phys. Rev.} B {\bf 77} 132501
\bibitem{Hol11} Holmqvist C, Teber S and Fogelström M  2011 {\it Phys. Rev.} B {\bf 83} 104521
\bibitem{Ber01} Bergeret F S, Volkov A F and Efetov K B 2001 {\it Phys. Rev. Lett.} {\bf 86} 3140
\bibitem{2Ber01} Bergeret F S, Volkov A F and Efetov K B 2001 {\it Phys. Rev. Lett.} {\bf 86} 4096
\bibitem{Ber05} Bergeret F S, Volkov A F and Efetov K B 2005 {\it Rev. Mod. Phys.} {\bf 77} 1321
\bibitem{Bla04} Blanter Y M and Hekking F W J 2004 {\it Phys. Rev.} B {\bf 69} 024525
\bibitem{Cro07} Crouzy B, Tollis S and Ivanov D A 2007 {\it
    Phys. Rev.} B {\bf 76} 134502
\bibitem{Bra07} Braude V and Nazarov Yu V 2007  {\it Phys. Rev. Lett.} \textbf{98} 077003
\bibitem{Bra08} Braude V and Blanter Ya M 2008 {\it Phys. Rev. Lett.} \textbf{100} 207001
\bibitem{Sos06} Sosnin I, Cho H, Petrashov V T and Volkov A F 2006 {\it Phys. Rev. Lett.} {\bf 96} 157002
\bibitem{Kei06} Keizer R S, Goennenwein S T B, Klapwijk T M, Miao G, Xiao G and Gupta A 2006 {\it Nature} (London) {\bf 439} 825
\bibitem{Kha10} Khaire T S, Khasawneh M A, Pratt W P Jr and Norman O. Birge 2010 {\it Phys. Rev. Lett.} 104 137002
\bibitem{Rob10} Robinson J W A, Witt J D S and Blamire M G 2010 {\it Science} {\bf 329} 5987
\bibitem{2Rob10} Robinson J W A, Halász G B, Buzdin A I and Blamire M G 2010 {\it Phys. Rev. Lett.} {\bf 104} 207001
\bibitem{Bry08} Brydon P M R, Kastening B, Morr D K and Manske D 2008 {\it Phys. Rev.} B {\bf 77} 104504
\bibitem{Bry09} Brydon P M R and Manske D 2009 {\it Phys. Rev. Lett.} {\bf 103} 147001
\bibitem{2Bry09} Brydon P M R 2009 {\it Phys. Rev.} B {\bf 80} 224520
\bibitem{3Bry09} Brydon P M R, Iniotakis C and Manske D 2009 {\it New Journal of Physics} {\bf 11} 055055
\bibitem{Naz94} Nazarov Y V 1994 {\it Phys. Rev. Lett.} {\bf 73} 1420
\bibitem{Naz99} Nazarov Y V 1999 {\it Superlattices and Microstructures} {\bf 25} 1221
\bibitem{bNaz05} Nazarov Y V 2005 {\it Quantum Transport and Circuit Theory} (Handbook
of Theoretical and Computational Nanotechnology, Micheal Rieth and
Wolfram Schommers, eds, American Scientific Publishers) Vol. 1
\bibitem{Cot09} Cottet A, Huertas-Hernando D, Belzig W and Nazarov Y V 2009 {\it Phys. Rev.} B {\bf 80} 184511
\bibitem{HH02} Huertas-Hernando D, Nazarov Y V and Belzig W 2002 {\it Phys. Rev. Lett.} {\bf 88} 047003
\bibitem{HH05} Huertas-Hernando D and Nazarov Y V 2005 {\it Eur. Phys. J.} B {\bf 44} 373
\bibitem{Shomali} Shomali Z, Zareyan M and Belzig W 2008 {\it Phys. Rev.} B {\bf 78} 214518
\bibitem{Bul77} Bulaevskii L N, Kuzii V V and Sobyanin A A 1977 {\it JETP Lett.} {\bf 25} 90
\bibitem{Buz82} Buzdin A I, Bulaevskii L N and Panyukov S V, 1982 {\it Pis'ma Zh. Eksp. Teor. Fiz.} {\bf 35} 147,{\it JETP Lett.} {\bf 35} 187
\bibitem{Buz91} Buzdin A I and Kupriyanov M Yu 1991 {\it Pis'ma Zh. Eksp. Teor. Fiz.} {\bf 53} 308, {\it JETP Lett.} {\bf 53} 321
\bibitem{Buz92} Buzdin A I, Bujicic B and Kupriyanov M Yu 1992 {\it Sov. Phys. JETP} {\bf 74} 124
\bibitem{Rya01} Ryazanov V V, Oboznov V A, Rusanov A Y, Veretennikov A V, Golubov A A and Aarts J 2001 {\it Phys. Rev. Lett.} {\bf 86} 2427
\bibitem{Cht01} Chtchelkatchev N M, Belzig W, Nazarov Y V and Bruder C 2001 {\it Pis'ma Zh. Eksp. Teor. Fiz.} {\bf 74} 357
\bibitem{Kon02} Kontos T, Aprili M, Lesueur J, Genêt F, Stephanidis B and Boursier R 2002 {\it Phys. Rev. Lett.} {\bf 89} 137007
\bibitem{Gui03} Guichard W, Aprili M, Bourgeois O, Kontos T, Lesueur J and Gandit P 2003 {\it Phys. Rev. Lett.} {\bf 90} 167001
\bibitem{Buz05} Buzdin A I 2005 {\it Rev. Mod. Phys.} {\bf 77} 935
\bibitem{Vas08} Vasenko A S, Golubov A A, Kupriyanov M Y and Weides M 2008 {\it Phys. Rev.} B {\bf 77} 134507
\bibitem{Sel03} Sellier H, Baraduc C, Lefloch F and Calemczuk R 2003 {\it Phys. Rev.} B {\bf 68} 054531
\bibitem{Moh06} Mohammadkhani G and Zareyan M 2006 {\it Phys. Rev.} B {\bf 73} 134503
\bibitem{Kon08} Konschelle F, Cayssol J and Buzdin A I 2008 {\it Phys. Rev.} B {\bf 78} 134505

\end{thebibliography}
\end{document}